\documentclass[10pt,onecolumn,aps,prl,amsmath,amssymb,notitlepage,nofootinbib,superscriptaddress]{revtex4}
\usepackage[latin9]{inputenc}
\usepackage{color}
\usepackage{verbatim}
\usepackage{amsmath}
\usepackage{amssymb}
\usepackage{graphicx}
\usepackage{esint}
\usepackage{epstopdf}

\makeatletter

\newcommand{\lyxdot}{.}

\@ifundefined{textcolor}{}
{%
 \definecolor{BLACK}{gray}{0}
 \definecolor{WHITE}{gray}{1}
 \definecolor{RED}{rgb}{1,0,0}
 \definecolor{GREEN}{rgb}{0,1,0}
 \definecolor{BLUE}{rgb}{0,0,1}
 \definecolor{CYAN}{cmyk}{1,0,0,0}
 \definecolor{MAGENTA}{cmyk}{0,1,0,0}
 \definecolor{YELLOW}{cmyk}{0,0,1,0}
}

\usepackage{epsfig}\usepackage{graphics}\usepackage{subeqnarray}\usepackage{amsfonts}
\providecommand{\U}[1]{\protect\rule{.1in}{.1in}}

\makeatother

\begin{document}

\title{Population aging through survival of the fit and stable}

\author{Tommaso Brotto}

\affiliation{Laboratoire de Physique Statistique de l'Ecole Normale Sup\'erieure,
CNRS UMR 8550-Universit\'e Paris 6-Universit\'e Paris 7; 24, rue
Lhomond, 75005 Paris, France}

\affiliation{Dipartimento di Fisica, Universit\'a degli Studi di Milano, Via
Celoria 16, 20133 Milano, Italy. INFN, Sezione di Milano, Via Celoria
16, 20133 Milano, Italy}

\author{Guy Bunin}

\affiliation{Department of Physics, Massachusetts Institute of Technology, Cambridge,
Massachusetts 02139, USA}

\author{Jorge Kurchan$^{1}$}
\begin{abstract}
Motivated by the wide range of known self-replicating systems, some
far from genetics, we study a system composed by individuals having
an internal dynamics with many possible states that are partially
stable, with varying mutation rates. Individuals reproduce and die
with a rate that is a property of each state, not necessarily related
to its stability, and the offspring is born on the parent's state.
The total population is limited by resources or space, as for example
in a chemostat or a Petri dish. Our aim is to show that mutation rate
and fitness become more correlated, \emph{even if they are completely
uncorrelated for an isolated individual}, underlining the fact that
the interaction induced by limitation of resources is by itself efficient
for generating collective effects. 
\end{abstract}
\maketitle
This work concerns the population dynamics of a system of self-replicating
individuals, with complex internal dynamics, where the state of the
system is not necessarily characterized by a genum, but could also
be the transcriptional situation. \textcolor{black}{A variety of scenarios
of self-replication have recently attracted considerable attention.
These include models of prebiotic evolution \cite{segre_prebiotic},
cells subjected to unforeseen challenges \cite{braun}, and artificial
self-replication and evolution \cite{artificial_self_rep}. In all
these cases, the complex dynamics might give rise to a wide spectrum
of replication cycle stabilities, implying `mutation rates' that show
large variability (here and in what follows `mutation rate' will be
used to denote changes even in non-genetic contexts)}. In fact, genetic
mutation rates themselves may vary by many orders of magnitude \cite{Saunders_phase_variation},
and epigenetic changes occur at yet another set of time-scales \cite{Schmitz_epigenetic_rate}.
Moreover, in systems that have not yet undergone a long process of
evolution, mutation rates might not be much smaller than growth rates.

The purpose of this paper is to study the effect that the spread of
mutation rates has on the population development, through its interplay
with fitness of the individuals, and to do so in a setting which assumes
as little as possible about the individuals' internal structure. Given
the wide range of disparate systems which are able to self-replicate
and subsequently evolve, it is natural to ask what phenomena will
be common to many of them. A `null model' for such a general situation
should include as little as possible of details of the systems' phase-space
(`phenotypes') and dynamics. For example, we wish to exclude effects
of long memory of the state in a \textit{single} individual after
many mutations, which may be highly relevant for genetics, and indeed
lead to effects such as genetic `hitchhiking' effects, or to the length
of adaptive walks, where an individual wanders on a given landscape
\cite{adaptive_walks}. In genetics, mutation rates are often low
compared to growth rates, and can often be taken to be constant or
assumed to have a small number of values. In contrast, the stability
of a dynamical self-replication state (interpreted as an inverse mutation
rate) may vary significantly between states, and cannot be taken a-priori
as much lower than growth rates.

To model the individuals' dynamics, trap models \cite{Bouchaud} have
been extensively used to give simple descriptions to highly complex
systems, including relaxation of glasses and rehology of soft and
biological materials \cite{trap_application1,trap_application2}.
Trap models have no memory of the state after leaving it, and the
new state is chosen at random. The model is therefore fully characterized
by the distribution of trap `stabilities', the average times spent
in a given trap. By considering a population of systems with can also
self-replicate, the resulting model belongs to the house-of-cards
class of models \cite{Kingman,Krug,Krug1}, but with a large dispersion
of mutation rates. The existence of multiple mutation rates has been
studied in the past \cite{Leigh,Ishii,Taddei,Lynch}. We are interested
here specifically in the phenomena derived from the combination of
\emph{(i)} a large number possible states (as one may envision in
a cellular system with more than a few phenotypic switches) and \emph{(ii)}
a large dispersion in the mutation rates. Individuals interact \emph{only}
via some constraint on the total population size, due to limitations
such as space or nutrients. Some of our results are, as we shall see,
remarkably insensitive to details.

\subsection*{Internal dynamics, selection.}

We shall assume that the system has $M(t)$ individuals, with an internal
dynamics that has states or attractors (for example a `cell fate')
labeled `\textbf{a}' that are not fully stable, mutations are random
with probability per unit time $\mu_{a}$. Being in a state \textbf{a}
confers an individual a rate of reproduction,\textcolor{black}{{} or
`fitness'} $\lambda_{a}$ per unit time. Therefore, each individual
is characterized by a pair of numbers $\left(\lambda,\mu\right)$.
The dynamics is as follows: when an individual leaves a state $\left(\lambda_{1},\mu_{1}\right)$,
it may fall on another state $(\lambda,\mu)$ with probability $p(\lambda,\mu)$,
which defines the distribution of states, and is independent of $\left(\lambda_{1},\mu_{1}\right)$
. After a replication, which is asexual, the two daughters inherit
the state of the parent. We will work in the limit where the number
of attractors $N$ is very large, so that the chances of falling twice
in the same one are negligible. This means that within this model
if two individuals have exactly the same $(\lambda,\mu)$, they are
in the same state as their common ancestor and have never mutated.
This model may be viewed as a `House of Cards' model with variable
mutation rates \cite{Kingman,Krug,Krug1}. Alternatively, it is a
`trap model' \cite{Bouchaud} with added self-replication.

Reproduction rates cannot be arbitrarily large because of physical
constraints, so the distribution must be bounded, $\lambda<\lambda_{max}$.
On the contrary, mutational timescales $1/\mu$ might be quite large,
as one may conceive that, for example, some phenotypic `cell fates'
may be very stable. We shall thus consider $\mu>\mu_{min}$, where
$\mu_{min}\ge0$ is always small, even zero. Although mutations and
fitness may be directly correlated, it is instructive and simpler
on a first approach to consider them independent; \emph{especially
because one of the main purposes of this paper is to pinpoint the
correlations that develop in a population but are absent for a single
individual}. We shall thus choose a product form 
\begin{equation}
p(\lambda,\mu)\propto Q(\lambda)P(\mu)\label{indep}
\end{equation}
$P(\mu)$ is defined for $\mu>\mu_{min}$, which we shall often take
as $0$. For small $\mu$ we shall consider below a power law ($P\sim\mu^{\alpha}$),
and more rapidly decaying distributions $P\sim e^{-\mu^{-n}}$ \cite{trap_model_comment}.
We shall assume the distribution of $\lambda$ falls to zero above
some $\lambda_{max}$, as $Q(\lambda)\sim(\lambda_{max}-\lambda)^{r}$,
with $\lambda\ge0$. The independence of the distributions corresponds
to the assumption of absence of a mechanism tailored to tune mutation
rates of an individual according to the external conditions. Another
possibility we could have considered is that the independent variables
are $\lambda$ and $\hat{\mu}\equiv\lambda/\mu$: 
\begin{equation}
\hat{p}(\lambda,\hat{\mu})\propto Q(\lambda)\bar{P}(\hat{\mu})\label{indep1}
\end{equation}
In order to complete the definition, it is natural to assume that
some mechanism, such as the total amount of food or space, constrains
the population. In both cases, our conclusion will be that {\em
a priori} uncorrelated fitness and mutation rates become correlated
through natural selection.

\section{Well-Mixed System}

Let us first consider a well-mixed situation, such as a chemostat.
Individuals compete for resources which limit their growth rate, and
also die or are removed from the system. Here we adopt the standard
Moran process, where population size is kept constant by removing
an individual at random every time a replication occurs.

\begin{figure}[ptb]
\begin{centering}
\includegraphics[clip,width=0.5\columnwidth]{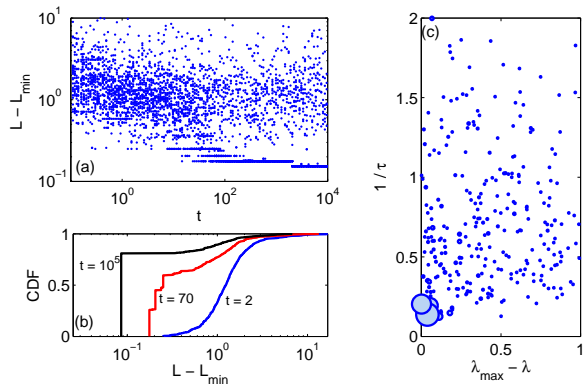} 
\par\end{centering}

\protect\protect\protect\caption{(a) Population evolution in time, stages (I)-(III), for $P\left(\mu\right)=\mu^{-2}e^{-1/\mu}$
and $M=1000$. Points are $L$-values of individuals. For clarity
1 in 40 individuals is plotted. (b) Cumulative distribution function
(CDF) of L-values, at $t=2,70,10^{5}$. (c) Distribution of states
in the $\left(\lambda,\mu\right)$-plane at $t=70$. Circle sizes
indicate the number of individuals in a state. Large clusters have
low $\mu$ and high $\lambda$, the continuous distribution less so.}

\label{clustering} 
\end{figure}

Starting from a population in a bounded distribution of $\lambda$
and $\mu$, it turns out one can distinguish four stages in the evolution,
see Fig. \ref{clustering}: (I) a \emph{continuous stage} in which
the population of all states is a negligible fraction of the total,
(II) a \emph{condensation stage} in which a significant fraction of
the population concentrates in a small range of values of \emph{`genetic
load' $L$ }defined as 
\begin{equation}
L\equiv\mu-\lambda\,.\label{eq:load_def}
\end{equation}
This leads to a (III) {\em takeover} phase characterized by rare
and rapid changes of single dominant states holding a finite fraction
of the population. These takeovers become statistically rarer as time
passes. This is analogous, for the well-mixed system, to the `Successional
Mutation Regime' of Ref. \cite{Fisher1}. Finally, there comes a time
when the average time between takeovers stabilizes because the system
has optimized as much as it can, this is the beginning of \emph{(IV)
saturation stage}. The population may, of course, start at one of
the later stages, and continue its evolution from there.

\subsection*{Individual and population stabilities}

In what follows we consider two types of time-scales, at the level
of the individual and the population. The individual's stability is
characterized by the time-scale $\tau\equiv1/\mu$, the typical time
to change its state. The rate of population change can be measured
by correlations of population composition, such as the overlap $C\left(t,t^{*}\right)$:
let $n_{a}\left(t\right)$ be the number of individuals in population
$i$ that are in state \textbf{a}, then $C\left(t,t^{*}\right)\equiv\sum_{a}min\left(n_{a}\left(t^{*}\right),n_{a}\left(t\right)\right)/M$,
where the sum runs over all states.

\begin{figure}[ptb]
\includegraphics[width=0.5\columnwidth]{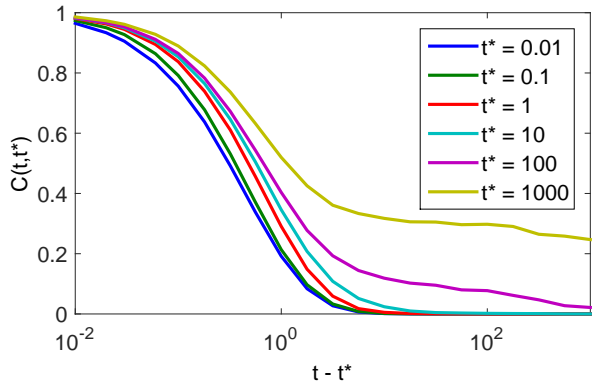}\protect\protect\caption{\textbf{Aging curves:} autocorrelation $C(t,t^{\ast})$ for $P\left(\mu\right)=\frac{2}{\sqrt{\pi}\mu^{2}}e^{-\mu^{-2}}$
in a system of size $M=1670$, averaged over 50 runs, showing that
evolution becomes progressively slower but never stops. The times
correspond to phases I to III.}

\label{aging1} 
\end{figure}

Below we find that in stages (I-III) the population \emph{ages}: time-scales
continue to grow as time passes, indicating long memory. This is illustrated
in Fig. \ref{aging1}. In stage (IV) aging is interrupted.

Let us now discuss the different stages in more detail, we consider
first $\mu_{min}=0$, and then we shall discuss the strictly positive
$\mu_{min}$ effects.

\subsection*{Stage (I), continuous distributions}

This stage lasts forever if the population $M$ is infinite. During
these times, the population of every type is negligible with respect
to $M$ i.e. $n_{a}\ll M\;\forall a$, so one may assume it to be
a continuous function of $\lambda,\mu$ and time $t$. Let $\rho\left(\lambda,\mu,t\right)$
be the normalized distribution of individuals in a population. It
satisfies 
\begin{equation}
\dot{\rho}=-(L+\langle\lambda\rangle)\rho+\langle\mu\rangle\;p(\lambda,\mu)\label{evolution}
\end{equation}
where $\langle A\rangle\equiv\int d\mu d\lambda\;A\;\rho(\lambda,\mu,t)$
and $L$ is the genetic load, Eq. (\ref{eq:load_def}).

First, we consider the stationary distribution $\rho_{s}(\lambda,\mu)$,
satisfying 
\begin{equation}
(-\mu+\lambda-\langle\lambda\rangle_{s})\rho_{s}+\langle\mu\rangle_{s}\;p(\lambda,\mu)=0\label{stationary}
\end{equation}
(here $\langle A\rangle_{s}$ denotes average over $\rho_{s}$). If
all the $\lambda$ are equal there is no selection pressure, and the
evolution (\ref{evolution}) converges in finite time to a stationary
distribution 
\begin{equation}
\rho_{stat}(\lambda,\mu)=\frac{p(\lambda,\mu)/\mu}{\int d\lambda d\mu\;p(\lambda,\mu)/\mu}\label{stationary1}
\end{equation}
which is simply the distribution weighted with the residence time
and always exists if the integral in the denominator is finite, which
we assume throughout. The individuals spend time in attractors with
finite lifetimes, making only rare visits to the more stable ones.

\begin{figure}
\includegraphics[width=0.5\columnwidth]{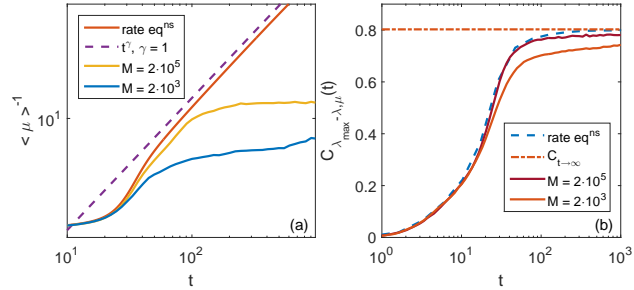}\protect\protect\caption{(a) Growth of individuals' stability $\left\langle \mu\right\rangle ^{-1}$
as a function of time, (b) Evolution of correlation $C_{\lambda_{max}-\lambda,\mu}\left(t\right)$,
between stability and fitness. $C_{t\rightarrow\infty}$ is the analytic
asymptotic result. Data for $P\left(\mu\right)=3\mu^{2}$, $\mu\le1$.}
\label{fig:tau_corr_growth} 
\end{figure}

Something dramatic happens as soon as we switch on a many-valued fitness:
the effect of \emph{any} fitness distribution is to switch the system
from short-lived to long-lived attractors. If $\mu_{min}=0$ the stationary
solution disappears altogether (Eq. (\ref{stationary}) has no solution,
see Appendix A). This is a\emph{ first order phase transition} to
a situation where the system forever evolves -- ages -- and population
time-scales grow indefinitely, see Fig. \ref{aging1}. The individuals'
stability also increases, spending time in states with larger and
larger time-scale $\tau=1/\mu$. For example, the increase of the
lifetime is given by: 
\begin{equation}
\langle\tau\rangle_{t}\ge\langle\mu\rangle_{t}^{-1}\sim t^{\gamma}
\end{equation}
where we have used Jensen's inequality, and the average is over the
distribution at time $t$, see Fig. \ref{fig:tau_corr_growth},(a).
The exponent $\gamma$ is one when $P(\mu)$ is a power law, and $\gamma=1/\left(n+1\right)$
for $P\sim e^{-\mu^{-n}}$ (see Appendix C). In this regime, there
is a clear selection pressure towards greater stability.

In order to quantify the correlations between stability and fitness,
we calculate the correlation coefficient $C_{\lambda_{max}-\lambda,\mu}\left(t\right)$
between $\mu$ and $\lambda_{max}-\lambda$. The asymptotic value
is always positive (see analytic expression Appendix D, Eq. (\ref{eq:C_tau_lambda})).
See Fig. \ref{fig:tau_corr_growth},(b) for Comparison between finite
$M$ simulations, rate equations (Eq. (\ref{evolution})), and analytical
asymptotics. Interestingly, models with strong memory (e.g. when fitness
jumps are small) can develop an opposite, negative correlation between
mutation and selection \cite{Gerrish}.

\subsection*{Stage (II), condensation}

Finite population size $M$ effects cannot be neglected, because they
can begin to show up at times of order $\ln M$, which for example
in chemostat conditions may be a few tens of generations \cite{braun}.
Two effects compete here: on the one hand for finite $M$ there is
an upper cutoff value $L{}_{c}=\mu_{c}-\lambda_{c}$, \textcolor{black}{set
by the probability of populating the tail with a single individual
}$\int_{\mu_{c}}^{\infty}\int_{\lambda_{c}}^{\lambda_{max}}d\mu d\lambda\rho(\lambda,\mu)\sim1/M$.
Secondly, there is the increase of the occupation of a few states
$n_{a}$, an effect already found in simpler models \cite{Kingman,Krug,Krug1,silvio},
analogous to the evolution of family names \cite{family_names}. Here,
this effect depends on $\mu$, and becomes especially strong for the
small $\mu$.

The net result is that after a time that is logarithmic in $M$ for
$P(\mu)\sim e^{-\mu^{-n}}$ and power law for power law tails at small
$P(\mu)$, the distribution begins to have clusters at population
in states of \emph{small $\mu$} (Fig. \ref{clustering}). At this
point the continuous description in terms of a $\rho(\lambda,\mu,t)$
breaks down. At the end of this crossover, a large fraction of the
population is in a single state having a $\mu_{d}$ of order $\ln M$
(or power law in $M$, for power law $P(\mu)$), while the rest of
the population is distributed in many states, most of which are less
stable and less fit.

\subsection*{Stage (III), Sweeps (Successional Mutation Regime)}

From the condensation time up to times exponential in $M$, the system
continues evolving by making rare and relatively rapid changes of
the dominant state (`sweeps'). Here a large number $M_{eff}^{a}$
individuals are in state $\left(\lambda_{a},\tau_{a}\right)$, but
there is a `cloud' of $M-M^{a}$ individuals which have suffered a
deleterious mutation to a state with substantially smaller $\lambda$,
constantly being renovated (see Figure \ref{clustering}). If the
cloud has $\lambda_{cloud}\ll\lambda_{a}$, one can easily see that
the dominant population size is 
\begin{equation}
M_{eff}^{a}=M\left(1-\frac{\mu_{a}}{\lambda_{a}}\right)\,.
\end{equation}

A sweep from a state $a$ to a state $b$ consists of a mutation in
a single individual from $a$ to $b$, followed by growth of the $b$
sub-population until fixation, where $a$-state individuals have gone
extinct. Mutations to lower $L=\mu-\lambda$ have a large probability
of fixation,$O\left(M^{0}\right)$. Mutations to higher $L$ rarely
fixate in the population, with probability exponentially small in
$M$.

In a simple case when the cloud size is negligible (i.e. $\mu_{a}/\lambda_{a}\ll1$
so $M_{eff}\simeq M$), the jump probability to $b$ in time $\delta t$
is well-known \cite{Nowak}, $\delta p_{a\rightarrow b}=\delta t\,\mu_{a}\left(1-r_{ab}\right)/\left(1-r_{ab}^{M}\right)$,
where $r_{ab}\equiv\lambda_{a}/\lambda_{b}$. For $\lambda_{b}>\lambda_{a}$
this is of order one even for large $M$, while for $\lambda_{b}<\lambda_{a}$
this is exponentially small in $M$, $\delta p_{a\rightarrow b}\sim\delta t\,\mu_{a}exp\left(-M\ln r_{ab}\right)$.

More generally, when a cloud is present and both $\lambda,\mu$ change
their value, it is still true that $p_{a\rightarrow b}=O\left(M^{0}\right)$
when $L_{b}<L_{a}$. For $L_{a}<L_{b}$

\begin{equation}
p_{a\rightarrow b}\sim\delta t\,\mu_{a}exp\left(-\frac{M_{eff}^{2}}{M}\ln r_{ab}\right)\,,\label{kimura}
\end{equation}
where now $r_{ab}\equiv L_{a}/L_{b}$, and $L_{a}=\mu_{a}-\lambda_{a}$
is the genetic load as before. This holds for small changes in both
$\lambda$ and $\mu$: $\left|\lambda_{b}/\lambda_{a}-1\right|,\left|\mu_{b}/\mu_{a}-1\right|\ll1$.
Here $M_{eff}\equiv M_{eff}^{a}\simeq M_{eff}^{b}$ (since changes
in $\lambda$,$\mu$ are small). The calculation is described in Appendix
E. Thus as in earlier stages, the genetic load $L_{a}=\mu_{a}-\lambda_{a}$
decreases as in a sequence of `record breaking' events \cite{record_stats},
except for random extinction events \cite{extinction} which are exponentially
suppressed in $M$.

In fact, as was noted before \cite{Berg_Lassig,Berg_Willmann_Lassig,Sella_Hirsh,Brotto},
Eq. (\ref{kimura}) implies the relation 
\begin{equation}
\frac{\delta p_{a\rightarrow b}}{\delta p_{b\rightarrow a}}\sim e^{\frac{M_{eff}^{2}}{M}(\ln L_{b}-\ln L_{a})}\,,\label{db1}
\end{equation}
which may be interpreted as detailed balance relation, {\em with
temperature $T=M/M_{eff}^{2}$ and energy}:

\begin{equation}
E_{a}=-\ln L_{a}\label{db2}
\end{equation}
The evolution may hence be seen, once the system is in the successional
mutation regime, as a relaxation within an energy landscape, Eq. (\ref{db2}),
in contact with a bath of inverse temperature $M_{eff}^{2}/M$. If
the jumps in $\lambda$ and $\mu$ are not small, detailed-balance
does not generally hold. 

In the above it was assumed that $Q(\lambda)$ has a bounded support
$\lambda<\lambda_{max}$, in which case the dynamics does is to rather
rapidly choose values of $\lambda_{a}\sim\lambda_{max}$, and decreasing
$\mu_{a}$. If, on the other hand, $Q(\lambda)$ is not bounded, the
situation is more subtle. One may still analyze the problem as an
annealing of $E(\lambda,\mu)$, with an `entropy' $S=\ln Q(\lambda)+\ln P(\mu)$.
In any case, if the tail of $Q(\lambda)$ falls fast enough, it is
always more convenient for the system to look for smaller $\mu$.

\subsection*{Stage (IV), saturation (`Interrupted Aging')}

Finally, at (often unobservable) large times, the maximal possible
improve in genetic load $\mu_{a}-\lambda_{a}-\mu_{max}+\lambda_{max}$
is such that the probability of a change decreasing genetic load becomes
comparable to that of an `extinction process' increasing it. At such
time aging stops, the system has reached its stationary regime. In
glassy physics this is known as `interrupted aging' \cite{Bouchaud}.
From the point of view of a system in contact with a bath of temperature
$T=M/M_{eff}^{2}$, the system has achieved thermal equilibrium 
\begin{equation}
p(a)\propto e^{-\frac{M_{eff}^{2}}{M}E_{a}}\propto e^{-\frac{M_{eff}^{2}}{M}\ln L_{a}}\label{db3}
\end{equation}
From this equation it is clear that for large $M$, small values of
$\mu$ are selected, via the attempt to reduce $L$.

\subsection*{Positive $\mu_{min}$: crossover to `condensation' dynamics. }

When $\mu_{min}>0$ is strictly positive but small, the aging process
we have described continues until $\langle\mu\rangle(t)$ decreases
to order of $\mu_{min}$. This may take very long, especially if the
distribution $P(\mu)$ falls fast, at which case the lower bound on
$\mu$ becomes irrelevant.

Let us consider first the case when $M$ is infinite. There is a critical
value $\lambda_{max}^{c}$: \emph{a)} If $\lambda_{max}<\lambda_{max}^{c}$,
Eq. (\ref{stationary}) has a smooth solution, the system has a stationary
distribution with many attractors having finite timescales, and starting
from any state, stationarity is reached in finite times. \emph{b)}
If $\lambda_{max}>\lambda_{c}$ the system has a stationary distribution
$\rho(\lambda,\mu)=\tilde{\rho}(\lambda,\mu)+a\delta(\mu-\mu_{min})\delta(\lambda-\lambda_{max})$
with a continuous part $\tilde{\rho}(\lambda,\mu)$ plus a finite
fraction $a$ of the population concentrated in $(\lambda_{max},\mu_{min})$,
see Appendix B for details. This phenomenon is closely analogous to
Bose-Einstein condensation in solid state physics, and has been encountered
previously in other evolving systems \cite{silvio}. In an experiment
starting from a random configuration, the system will evolve towards
this distribution, but condensation of a finite fraction in $(\lambda_{max},\mu_{min})$
necessarily takes times divergent with $M$. As before, the population
ages, occupying states with $\left(\lambda,\mu\right)$-values which
are increasingly closer to $(\lambda_{max},\mu_{min})$. An example
of a phase-diagram is given in Appendix B. This transition is analogous
to Eigen's mutational meltdown \cite{eigen}, where evolution is hampered
by high mutation rates.

If on the other hand both $M$ and $\mu_{min}$ are finite, the system
has a complex behavior of competition between the condensation effect
due to finiteness of $\mu_{min}$ at infinite $M$ and those due to
finiteness of $M$ (i.e. equilibration, stages \emph{(II)} and \emph{(III)}).

\subsection*{Relation between fitness and mutation rate in the well-mixed case}

Let us now summarize how the individuals' stability $\tau=1/\mu$
is selected in the different stages.

In the continuous \emph{Stage I} the average of $\tau$ grows steadily
as $\langle\tau\rangle\ge\langle\mu\rangle^{-1}\sim t^{\gamma}$ where
$\gamma$ depends on the distributions.

Once in \emph{Stage III}, the dominant type changes in rather fast
sweeps. Mutations to lower $L=\mu-\lambda$ have a significant probability
of fixation, while fixations to higher $L$ are rare. Therefore mutation
rates decrease as growth rates increase. If $P(\lambda)$ has a bounded
support $\lambda<\lambda_{max}$, what the dynamics does is to rather
rapidly choose values of $\lambda_{i}\sim\lambda_{max}$, and increasing
$\tau_{i}$. If, on the other hand, $P(\lambda)$ is not bounded,
the situation is more subtle. One may analyze the problem as an annealing
of $E(\lambda,\mu)$, see Eq. (\ref{db2}) with an `entropy' $S=\ln Q(\lambda)+\ln P(\mu)$.
In any case, if the tail of $Q(\lambda)$ falls fast enough, it is
always more convenient for the system to look for smaller $\mu$.

Finally, phase (IV), corresponding to equilibrium, may be analyzed
directly on the basis of Eq (\ref{db3}). It is clear that in this
final value large values of $\mu$ are selected.

\section{Expansion in space}

Let us turn to a situation where the limited resource is not nutrient,
but rather space. We model spatial expansion in two and three dimensions
with cells growing radially or confined between walls as follows:
cells are modeled as circles (or spheres in three-dimensions) of equal
radius. They attempt to reproduce with a rate $\lambda$. The offspring
(having the same $\lambda$ and $\mu$), is created in contact with
the parent cell - if there is no free place in contact with the mother
cell reproduction does not happen. Just as in the previous section,
cells mutate to with their rate $\mu$ to a new state with $(\lambda^{\prime},\mu^{\prime})$
chosen with probability $Q(\lambda^{\prime})P(\mu^{\prime})$. Cells
in the bulk of the colony produce no more progeny, but they keep on
mutating.

Here we are interested in the following question: what is the effect
of a dispersion in the values of mutation rate $\mu$? Unlike the
case of a well-mixed system, where the `genetic load' (the rate loss
of population of a state) was naturally $L_{a}=\mu_{a}-\lambda_{a}$,
and this led naturally to a decrease in $\mu$ in the population,
here it is not clear if this should happen at all. As it turns out,
\emph{even in this spatial version smaller $\mu$ are selected, but
the way this happens is less obvious.}

\begin{figure}[ptb]
\includegraphics[width=0.5\columnwidth]{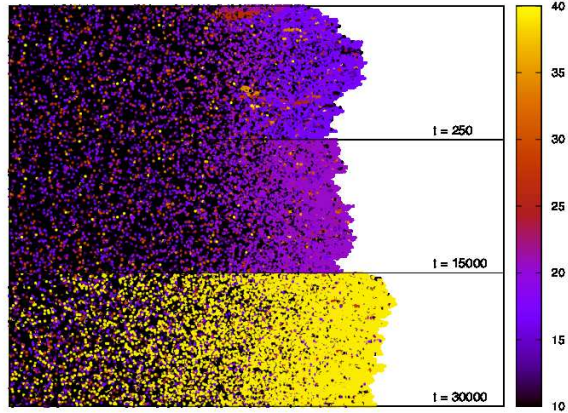}\protect\protect\protect\caption{Population of cells growing between two walls (along x direction,
left to right) with a distribution of $(\lambda,\mu)$ given by $Q(\lambda)P(\mu)$.
The figure shows three snapshots of the more advanced $\Delta G$
of the growing colony, at times $t=250,15000,30000$. The color code
refers to the values of $1/\mu$ of particles. The front layer of
the colony is composed of cells selected to have unusually low mutation
rate, and consistently, this layer becomes thicker with time.}

\label{linear} 
\end{figure}

In order to understand the basic mechanism, following Refs. \cite{KuhrFrey,Lavrentovich},
we first consider cells that may be in only two states, with $\lambda_{f}>\lambda_{s}$.
We consider the growth of a linear front between two walls (Fig. \ref{linear}).
The dominating `fast' type mutates into the `slow' type with typical
time-scale $1/\mu$, which we assume is large so that a large colony
is essentially composed of `fit' cells with $\lambda_{f}$, with occasional
`spots' of cells that where born `unfit', with $\lambda_{s}$. (A
complete extinction of the cells with $\lambda_{f}$ would be irreversible,
but its probability vanishes with the size if $\lambda_{f}>\lambda_{s}$).

\begin{figure}[ptb]
\includegraphics[angle=270,width=0.5\columnwidth]{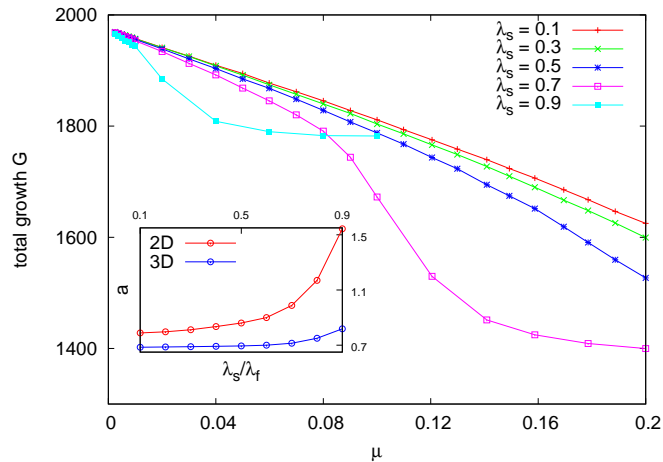}\protect\protect\protect\caption{Values of the total linear growth $G$ for a clone with two states
$\lambda_{f}$ and $\lambda_{s}$. The plots are for $\lambda_{f}=1$
and different values of mutation rates $\mu$ and $\lambda_{s}$.
The is linear regime for small $\mu$ (independent `spots') with gradient
$-A(\lambda_{s})$, which is the one described by equation (\ref{eqfactor}).
Surprisingly enough, growth is more strongly retarded for higher values
of $\lambda_{s}$, meaning that deleterious mutations are worse when
they are not completely sterile. (In the curves for $\lambda_{s}=0.7$
and $0.9$, one can see the transition to the inactive phase, signaled
in the steep drop in the values of the colony growth for high $\mu$
values.) }

\label{crescitavs1sutau} 
\end{figure}

The results for the colony length $G$ are shown in Fig. \ref{crescitavs1sutau}:
the growth rate of a $\lambda_{f}$-rich population is indeed slowed
down by the mutations. The reason is simple: a mutation creates a
`spot' of slowly reproducing cells with $\lambda_{s}$, which becomes
an obstacle and delays the advance of the from of cells with $\lambda_{f}$.
If $\lambda_{s}=0$, the mutated cells leave no offspring, but for
larger values of $\lambda_{s}$ the `bad spots' become larger -- although
less inefficient. In order to quantify the interplay between fitness
and stability, consider the case when $\tau$ is so large that two
different spots due to different mutations do not overlap. The retardation
factor $R$ on the growth length $G$, with respect to $G_{\infty}$
-- the one that would be obtained without deleterious mutations --
is then proportional to the density of spots, in turn $\propto\mu$,
and we get: 
\begin{align}
R\equiv\frac{G_{\infty}-G}{G_{\infty}}=A(\lambda_{s}/\lambda_{f})\;\frac{\mu}{\lambda_{f}}\label{eqfactor}
\end{align}
The function $A(\lambda_{s}/\lambda_{f})$ is the retardation per
unit `spot', which depends on the dimension and on the details of
the dynamics. We have computed $G$ numerically in two and three dimensions,
and we find that (see Fig. \ref{crescitavs1sutau}) $A(\lambda_{s}/\lambda_{f})$
is a \emph{growing} function of $\lambda_{s}/\lambda_{f}$: {\em
The net counter intuitive result is that `bad' mutations are more
deleterious if they are just barely worse, $\lambda_{s}\lesssim\lambda_{f}$,
while they are more innocent if they are completely sterile ($\lambda_{s}=0$).}

Consider next two spatially adjacent colonies with only two states
as above, $(\lambda_{f}^{1},\lambda_{s}^{1},\mu^{1})$, $(\lambda_{f}^{2},\lambda_{s}^{2},\mu^{2})$,
respectively. The case of two competing colonies with ($\lambda_{1}$,$\lambda_{2}$)
and no mutations allowed ($\mu_{1}=\mu_{2}=0$), has been studied
extensively by Korolev et al. \cite{HallaNelson}. The colony with
the highest $\lambda$ prevails, the time for the overcome scales
linearly in $\Delta\lambda$ in the case of colony growing with a
linear front, while it scales logarithmically in a radial, two-dimensional
growth. Here, the relevant parameters for the competition of the colonies
are not $\lambda_{f}^{1}$ and $\lambda_{f}^{2}$, but rather their
net growth rate $\lambda_{f}^{1}(1-R_{1})$ and $\lambda_{f}^{2}(1-R_{2})$,
which is affected by their respective `retardation' factors of $R_{1}$
and $R_{2}$ calculated as in Eq. (\ref{eqfactor}).

\begin{figure}[ptb]
\includegraphics[width=0.5\columnwidth]{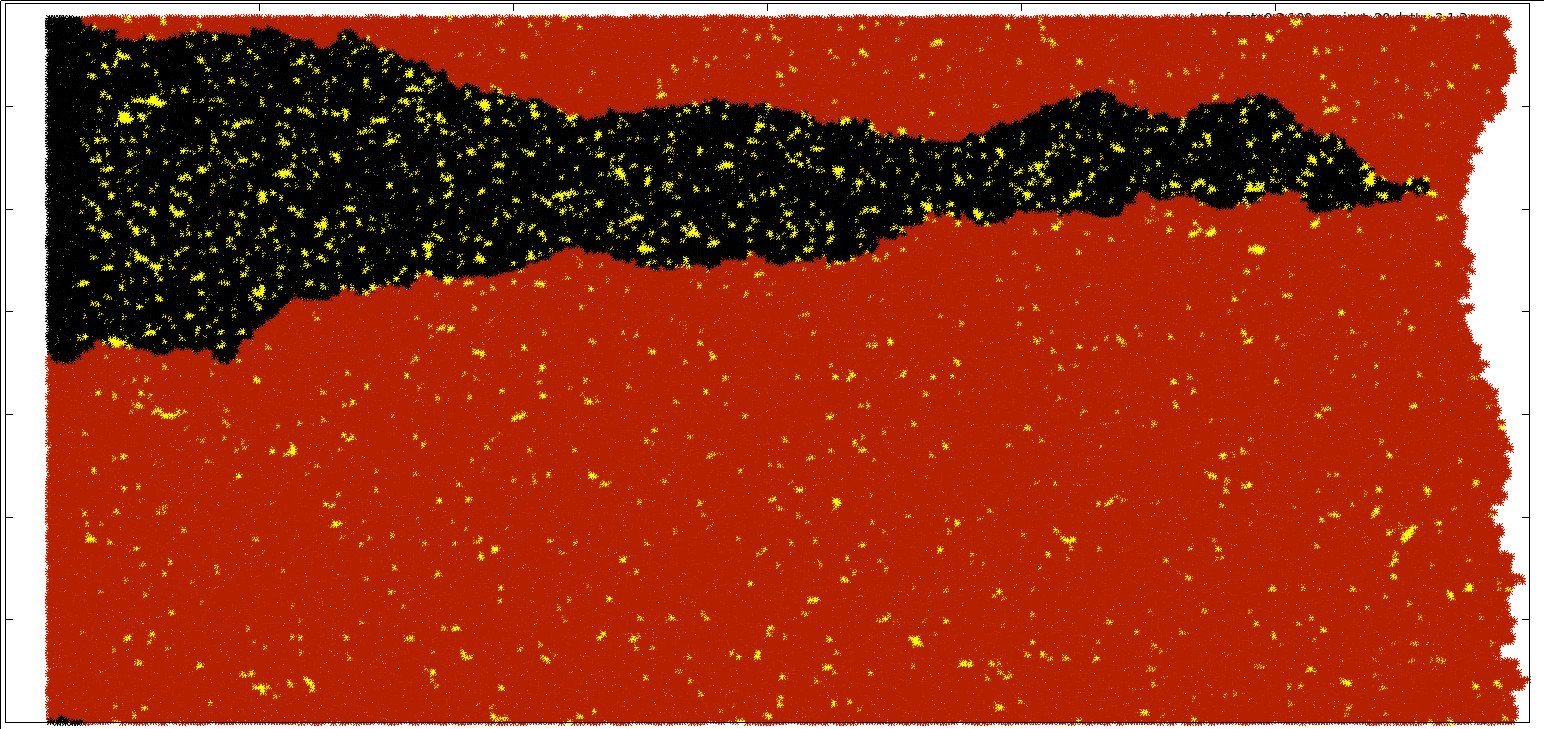}\protect\protect\protect\caption{Competition between two species of cells with equal $\lambda_{f}$,
mutating into equal $\lambda_{s}=0.3\lambda_{f}$, but with two different
mutation rates. The first species (black) has $\mu_{1}=1/20$, while
the second one (red) has $\mu_{2}=1/100$. Particles which mutated
while in the front are shown in yellow. At the beginning (far left)
the two species have the same number of cells. The second species
grows faster and overtakes the first one. This can be seen also by
looking at the front shape after the overcoming: the front to the
right of the black species is retarded. The effect is only due to
a difference in the mutation rate $\tau$: the colony composed of
stabler cells prevails.}

\label{taucompetition} 
\end{figure}

Fig. \ref{taucompetition} shows two competing sets with the same
fitness $\lambda_{f}^{1}=\lambda_{f}^{2}$, $\lambda_{s}^{1}=\lambda_{s}^{2}$,
but different values of $\mu$, and hence different $R$'s. The type
with the smallest $\tau$ becomes extinct, because of the slowing
down provoked by the occasional mutations, which are more frequent
in one case than in the other.

In a spatial setting it is important to notice that the relevant dynamics
and the effects of selection are present only at the advancing front
of the colony. In the bulk, where cells have no more space to reproduce,
if we assume they may continue to mutate they eventually go to equilibrium,
i.e. the bulk population eventually (in a finite time) will sample
the original probability distribution for $\lambda$ and $\mu$, therefore
losing any evolutionary achievement the system had reached.

We are now in a position to return to the original model, with a distribution
of values of $\lambda$ and $\mu$. The sequence of phases described
in the well-mixed case is present, with the same features, also in
the spatial setting with competing colonies, as already discussed
by Desai and Fisher \cite{Fisher1}. At late stages, when the evolution
is essentially successional, the situation is quite similar to the
one discussed in the simplified models above. There is a dominant
type with $\lambda_{f}$ close to the optimum, and there will be occasional
mutations, most of which will produce less fit cells. This yield unsuccessful
`spots' of low fitness, that retard the advance, just as in the simpler
example above. The only difference with the `two-type' version is
that the value of $\lambda_{s}$ will not be fixed, but taken from
the $Q(\lambda)$ distribution, which has the only effect of modifying
the function $A$ in Eq. (\ref{eqfactor}) to a $Q$-dependent function
$A_{Q}$. When a mutation finally produces a lineage with smaller
$L=A_{Q}\mu-\lambda$, it will with high probability overcome the
dominant one, even if it has a lower $\lambda_{f}$.

All in all we find the non-trivial geometric factor $A_{Q}$ which
modifies the genetic load with respect to the well-mixed case, but
otherwise produces the same selection of large $\mu$'s effect. $A_{Q}$
turns out to be greater in two spatial dimensions than in 3D, and
it is of order one. In Fig. \ref{linear} we see how selection comes
about in a population as the one described for the well-mixed case:
the cells at the front are the most stable ones.

\section{Conclusions}

The stability and reproduction rates become correlated for a population
even if there was no such mechanism for the single individual to start
with. If one considers the statistics at long times, a large part
of the population is in very fit (high $\lambda$) and very stable
(low $\mu$) states, while the rest of the population is in states
that have intermediate values of both $\lambda$ and $\mu$: the fittest
tend to be the stablest individuals. This conclusion seems to be valid
whatever the mechanism that limits the population. If we observed
an internal (e.g. chemical) mechanism that is responsible for the
higher stability of some cells, we might conclude that it is a well-designed
response of an individual cell tailored to lower its mutation rate
under favorable conditions. It could be, however, that as in our case
there is no such mechanism, and the correlation only arises at the
level of population, for purely statistical reasons. In order to test
whether an adaptive mechanism that changes mutation rate in response
to external conditions exist -- that is to say: whether fitness and
mutation rates correlate for the states of \emph{a single cell} --
we could try following the changes of a single individual, ignoring
its progeny. Another relation between fitness and mutation rate appears
when the environment changes: changes in the fitness of states $\lambda_{a}$
amount to `rejuvenating' (reinitializing) the system: this brings
about a drop in both the fitness and the average mutation time (see
\cite{Leibler}).

`Population Aging' in which the dynamics slows down arises also as
a purely collective effect (see Ref. \cite{Krug}), with the added
element that the stability of the individuals themselves also increases.
The slowing down of the dynamics may be checked by means of an experiment
to test the divergence time $t_{div}$ of subpopulations isolated
at different $t^{*}$.

\subsection*{Acknowledgments}

We wish to thank A. Amir, E. Braun, N. Brenner and L. Peliti for useful
suggestions. G.B. acknowledges the support of the Chateaubriand Fellowship
and the Pappalardo Fellowship in Physics. T.B. acknowledges the support
of UIF/UFI (Bando Vinci).

\onecolumngrid

\section*{Appendices}

\subsection*{Appendix A: The stationary solution, and the $\mu_{min}=0$ case}

Denote $R_{\mu}\equiv\left\langle \mu\right\rangle \equiv\int d\mu d\lambda\mu\rho\left(\lambda,\mu\right)$,
and $R_{\lambda}\equiv\left\langle \lambda\right\rangle =\int d\mu d\lambda\rho\left(\lambda,\mu\right)\lambda$.

$p(\lambda,\mu)$ is defined on $\lambda_{\min}<\lambda<\lambda_{max}=1$
and $0\le\mu_{min}<\mu<\mu_{max}$. The condition for stationarity
is 
\begin{equation}
\dot{\rho}_{s}=\left[\lambda-\mu-R_{\lambda}\right]\rho_{s}+R_{\mu}p(\lambda,\mu)=0\ ,\label{eq:stationary}
\end{equation}
which gives the stationary $\rho_{s}\left(\lambda,\mu\right)$ 
\begin{equation}
\rho_{s}(\lambda,\mu)=\frac{R_{\mu}p(\lambda,\mu)}{\mu-\lambda+R_{\lambda}}\ .
\end{equation}
We look for the conditions under which $\rho_{s}\left(\lambda,\mu\right)$
exists, and is a smooth function that is not everywhere zero. The
last condition requires $R_{\mu}>0$.

If $R_{\mu}=0$, no normalizable steady-state solution exists. As
we show in the next section, the system will age.

Assuming these conditions are met, we obtain
\begin{align}
1 & =R_{\mu}\int d\lambda d\mu\;\frac{p(\lambda,\mu)}{\mu-\lambda+R_{\lambda}}\ ,\label{eqs1_a}\\
R_{\mu} & =R_{\mu}\int d\lambda d\mu\;\frac{\mu\;p(\lambda,\mu)}{\mu-\lambda+R_{\lambda}}\ ,\label{eqs1_b}\\
R_{\lambda} & =R_{\mu}\int d\lambda d\mu\;\frac{\lambda\;p(\lambda,\mu)}{\mu-\lambda+R_{\lambda}}\ .\label{eqs1_c}
\end{align}
Of the three equations only two are independent\footnote{Eq. (\ref{eqs1_c}) minus Eq. (\ref{eqs1_b}) minus $R_{\lambda}$
times Eq. (\ref{eqs1_a}) gives an identity.}.

From Eq. (\ref{eqs1_b}), if $R_{\mu}>0$ then 
\begin{equation}
1=\int d\lambda d\mu\frac{\mu\;p(\lambda,\mu)}{\mu-\lambda+R_{\lambda}}\ ,\label{eq2}
\end{equation}
which fixes $R_{\lambda}$. In that case, $R_{\mu}$ is obtained from
the Eq. (\ref{eqs1_c}).\\

We note the following:

$\bullet$ For $R_{\mu}>0$, the integral in Eq. (\ref{eqs1_a}) must
be finite, so the integral must converge which requires $R_{\lambda}\geq\lambda_{max}-\mu_{min}$.
We also have $R_{\lambda}=\left\langle \lambda\right\rangle \leq\lambda_{max}$.
Therefore, if $\mu_{min}=0$, we find $R_{\lambda}=\lambda_{max}$.
But then Eq. (\ref{eq2}) cannot be satisfied for general $p(\lambda,\mu)$.
We conclude that if $\mu_{min}=0$ then $R_{\mu}=0$ and $R_{\lambda}=\lambda_{max}$.

$\bullet$ If $\mu_{min}>0$ there may be a solution to equation (\ref{eq2})
with $R_{\lambda}\neq\lambda_{max}$. Note that \emph{if $\lambda_{\min}=\lambda_{\max}$
Eq. (\ref{eq2}) is automatically satisfied, so there is always a
smooth solution. Hence, for }$\mu_{min}=0$\emph{ the system ages
as soon as there is any dispersion in the possible values of }$\lambda$.

\subsection*{Appendix B: $\mu_{min}>0$ condensation}

\begin{figure}[ptb]
\includegraphics[width=0.4\columnwidth]{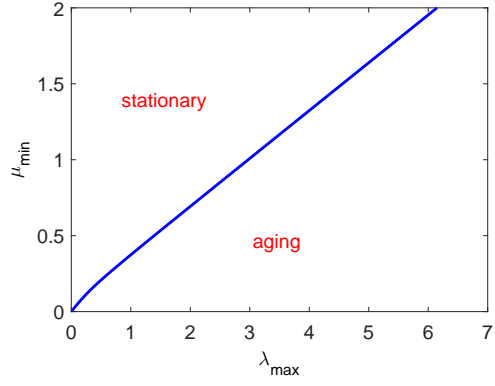}

\protect\protect\protect\caption{The phase diagram in terms of $\lambda_{max}$ and $\mu_{min}$, for
$p\left(\mu\right)\propto\mu^{-2}e^{-1/\mu}$ in $\left[\mu_{min},\infty\right]$.
Similar diagrams arise in glassy models with metastable states \cite{garrahan}. }

\label{diagram} 
\end{figure}

If $\mu_{min}>0$ and there is no smooth solution, we must extend
the solution space. We attempt the following $\rho=\rho_{smooth}+a\delta(\lambda-\lambda_{max})\delta(\mu-\mu_{min})$,
meaning that a condensate forms at a single trap at $\left(\lambda_{max},\mu_{min}\right)$.
Let $D_{\varepsilon}$ be a small region in $\left(\lambda,\mu\right)$-space
that contains the point $\left(\lambda_{max},\mu_{min}\right)$. Integrating
the stationary condition, Eq. (\ref{eq:stationary}), over $D_{\varepsilon}$
we find 
\[
a\cdot\left(\lambda_{max}-\mu_{min}-R_{\lambda}\right)+\int_{D_{\varepsilon}}d\lambda d\mu\left[\left(\lambda-\mu-R_{\lambda}\right)\rho_{smooth}+R_{\mu}p(\lambda,\mu)\right]=0\ ,
\]
and as the area of $D_{\varepsilon}$ goes to zero, this approaches
$a\cdot\left(\lambda_{max}-\mu_{min}-R_{\lambda}\right)=0$, so that
$R_{\lambda}=\lambda_{max}-\mu_{min}$ for $a\neq0$. The two independent
equations (\ref{eqs1_a},\ref{eqs1_b}) now read 
\begin{align}
1 & =a+R_{\mu}\int d\lambda d\mu\frac{p(\lambda,\mu)}{\mu-\lambda+R_{\lambda}}\ ,\nonumber \\
R_{\mu} & =a\mu_{min}+R_{\mu}\int d\lambda d\mu\frac{\mu\ p(\lambda,\mu)}{\mu-\lambda+R_{\lambda}}\ ,\label{eqs1a}
\end{align}
and obtain 
\begin{equation}
a=\frac{1-\int d\lambda d\mu\frac{\mu\ p(\lambda,\mu)}{\mu-\lambda+R_{\lambda}}}{1-\int d\lambda d\mu\frac{p(\lambda,\mu)}{R_{\lambda}-\lambda+\mu}\left(\mu-\mu_{min}\right)}\ ,\label{eq2a}
\end{equation}
which using $R_{\lambda}=\lambda_{max}-\mu_{min}$ fixes $a$. In
Fig. \ref{diagram}, an example of a phase diagram is shown, for $p\left(\mu\right)\propto\mu^{-2}e^{-1/\mu}$.
Below the line, a solution with $a>0$ exists; Above the line, a solution
to Eq. (\ref{eq:stationary}) exists.

\subsection*{Appendix C: Aging for the $M=\infty$ case\label{sub:Aging-for-the_inifite}}

In the following, to make expressions simpler we take $\lambda_{max}=1$,
which can always be obtained by rescaling time. Denote $u\equiv1-\lambda$.
The time-dependent solution for $\rho$ is 
\[
\rho=p\left(\lambda,\mu\right)\int_{0}^{t}dt^{\prime}e^{-\left(u+\mu\right)\left(t-t^{\prime}\right)}e^{\int_{t^{\prime}}^{t}R_{1-\lambda}\left(t^{\prime}\right)dt^{\prime}}R_{\mu}\left(t^{\prime}\right)+\rho_{0}e^{-\left[\left(u+\mu\right)t+\int_{0}^{t}R_{1-\lambda}\left(t^{\prime}\right)dt^{\prime}\right]}\ .
\]
The time-dependent equivalents of Eq. (\ref{eqs1_a},\ref{eqs1_b},\ref{eqs1_c})
read 
\begin{align}
1 & =e^{\int_{0}^{t}R_{1-\lambda}(t^{\prime\prime})dt"}\int_{0}^{t}dt^{\prime}f(t^{\prime})\;I(t-t^{\prime})+const\cdot e^{-\int_{0}^{t}R_{1-\lambda}(t^{\prime\prime})dt"}\nonumber \\
R_{\mu} & =e^{\int_{0}^{t}R_{1-\lambda}(t^{\prime\prime})dt"}\int_{0}^{t}dt^{\prime}f(t^{\prime})\;I_{\mu}(t-t^{\prime})+\;const^{\prime}e^{-\int_{0}^{t}R_{1-\lambda}(t^{\prime\prime})dt"}\nonumber \\
R_{1-\lambda} & =e^{\int_{0}^{t}R_{1-\lambda}(t^{\prime\prime})dt"}\int_{0}^{t}dt^{\prime}f(t^{\prime})\;I_{1-\lambda}(t-t^{\prime})+const^{\prime\prime}e^{-\int_{0}^{t}R_{1-\lambda}(t^{\prime\prime})dt"}\label{eqs1_dynam}
\end{align}
the functions $I\left(t\right),I_{\mu}\left(t\right),I_{1-\lambda}\left(t\right)$
are defined as 
\begin{align}
I\left(t\right) & \equiv\int d\mu d\lambda p\left(\lambda,\mu\right)e^{-\left(u+\mu\right)t}\nonumber \\
I_{\mu}\left(t\right) & \equiv\int d\mu d\lambda\mu p\left(\lambda,\mu\right)e^{-\left(u+\mu\right)t}\nonumber \\
I_{1-\lambda}\left(t\right) & \equiv\int d\mu d\lambda\left(1-\lambda\right)p\left(\lambda,\mu\right)e^{-\left(u+\mu\right)t}
\end{align}
and 
\begin{equation}
f(t)\equiv R_{\mu}(t)e^{-\int_{0}^{t}R_{1-\lambda}(t^{\prime\prime})dt"}\ .
\end{equation}

We are now interested in the large time behavior of $R_{\mu}$ and
$R_{1-\lambda}\equiv\left\langle 1-\lambda\right\rangle $. (As before,
we take $\lambda_{max}=1$.) If the initial condition has support
only over $\mu>0$, the result does not depend on the last term in
the three equations for large times. We shall use the following result:
if both $f$ and $I_{\ast}$ decrease as a power-law or exponent,
then for large times 
\begin{equation}
\int_{0}^{t}\;f(t^{\prime})I_{\ast}(t-t^{\prime})dt^{\prime}\sim f(t)\;\left[\int_{0}^{\infty}dt^{\prime}\;I_{\ast}(t^{\prime})\right]+I_{\ast}(t)\;\left[\int_{0}^{\infty}dt^{\prime}\;f(t^{\prime})\right]
\end{equation}
Unless both fall equally fast, only one of these two terms contribute.
We shall assume (and later check) that $I(t)$ falls \emph{slower}
than $f(t)$. Then, the first of Eq. (\ref{eqs1_dynam}) becomes 
\begin{equation}
1=e^{\int_{0}^{t}R_{1-\lambda}(t^{\prime\prime})dt"}\;I(t)\;\left[\int_{0}^{\infty}dt^{\prime}\;f(t^{\prime})\right]
\end{equation}
which means that 
\begin{equation}
R_{1-\lambda}(t)\sim-\frac{\partial\ln I(t)}{\partial t}\label{aa}
\end{equation}
Next, we shall assume that $I_{\mu}$ fall \emph{as fast as} $f(t)$.
We have that the second of Eq. (\ref{eqs1_dynam}) reads: 
\begin{equation}
R_{\mu}=e^{\int_{0}^{t}R_{1-\lambda}(t^{\prime\prime})dt"}\left\{ f(t)\;\left[\int_{0}^{\infty}dt^{\prime}\;I_{\mu}(t^{\prime})\right]+I_{\mu}(t)\;\left[\int_{0}^{\infty}dt^{\prime}\;f(t^{\prime})\right]\right\} 
\end{equation}
Dividing by the expression for $f$ we find 
\begin{equation}
R_{\mu}\propto\left\{ \frac{f(t)}{I(t)}\;\left[\int_{0}^{\infty}dt^{\prime}\;I_{\mu}(t^{\prime})\right]+\frac{I_{\mu}(t)}{I(t)}\;\left[\int_{0}^{\infty}dt^{\prime}\;f(t^{\prime})\right]\right\} 
\end{equation}
which by assumption grows as $\frac{I_{\mu}(t)}{I(t)}$ does. Finally,
$I_{1-\lambda}$ falls faster than both $f$ and $I_{\mu}$ so that:
\begin{equation}
R_{1-\lambda}=e^{\int_{0}^{t}R_{1-\lambda}(t^{\prime\prime})dt"}f(t)\;\left[\int_{0}^{\infty}dt^{\prime}\;I_{1-\lambda}(t)\right]\propto R_{\mu}(t)\label{bb}
\end{equation}
where we have used the previous expression for $f$.

Let us now give some examples. Performing the integral over $\lambda$,
we get: 
\begin{align}
I(t) & =\frac{1-e^{-t}}{t}\int d\mu\;p(\mu)e^{-\mu t}\nonumber \\
I_{\mu}(t) & =\frac{1-e^{-t}}{t}\int d\mu\;\mu p(\mu)e^{-\mu t}\nonumber \\
I_{1-\lambda}(t) & =\frac{1-e^{-t}(1-t)}{t^{2}}\int d\mu\;p(\mu)e^{-\mu t}
\end{align}

$\bullet$ \textbf{Power law }$P(\mu)$. If $P(\mu)\propto\mu^{\alpha}$,
and $Q\left(\lambda\right)=1$ on $\lambda\in\left[0,1\right]$: We
have that $\int d\mu\;p(\mu)e^{-\mu t}\sim t^{-\alpha+1}$ and $\int d\mu\;\mu p(\mu)e^{-\mu t}\sim t^{-\alpha}$,
so that for large $t$ we get: 
\begin{align}
I(t) & \sim t^{-\alpha}\nonumber \\
I_{\mu}(t) & \sim t^{-\alpha-1}\nonumber \\
I_{1-\lambda}(t) & \sim\frac{1}{t}I(t)
\end{align}
Equations (\ref{aa}) and (\ref{bb}) give: 
\begin{align}
R_{1-\lambda}(t) & \sim-\frac{\alpha}{t}\\
R_{\mu}(t) & \propto t^{-1}\label{eq:R_tau_powerlaw}
\end{align}

$\bullet$ \textbf{Strongly suppressed }$P(\mu)$.\textbf{ }If $P(\mu)\propto e^{-b\mu^{-n}}$,
and $Q\left(\lambda\right)=1$ on $\lambda\in\left[0,1\right]$: The
integrals over $P(\mu)$ may be evaluated by saddle point. The saddle
point is $\mu_{sp}=\left(\frac{t}{bn}\right)^{-\frac{1}{n+1}}$. We
get 
\begin{equation}
I(t)=e^{\left(\frac{t}{t_{0}}\right)^{-\frac{n}{n+1}}}\;\;\;;\;\;\;\frac{I_{\mu}}{I}=\mu_{sp}=\left(\frac{t}{bn}\right)^{-\frac{1}{n+1}}
\end{equation}
and: 
\begin{align}
R_{1-\lambda}(t) & \sim t^{-\frac{1}{n+1}}\\
R_{\mu}(t) & \propto t^{-\frac{1}{n+1}}\label{eq:R_tau_exp_Gauss}
\end{align}

$\bullet$ \textbf{Other }$Q\left(\lambda\right)$.\textbf{ }The above
results are unchanged if more generally, $P\left(\lambda\right)=\left(1-\lambda\right)^{m}$
with $m>0$, for $\lambda$ around $1$. The power of $I(t)$ changes,
but $R_{1-\lambda}(t)\sim-\frac{\partial\ln I(t)}{\partial t}$ is
unchanged, and $R_{1-\lambda}\sim R_{\mu}(t)$.

In all cases one may check that these asymptotics are consistent with
the initial assumptions.

\subsection*{Appendix D: Correlations between fitness and stability}

To study correlations we introduce 
\begin{equation}
R_{\mu,1-\lambda}\equiv\left\langle \left(1-\lambda\right)\mu\right\rangle =e^{\int_{0}^{t}R_{1-\lambda}(t^{\prime\prime})dt"}\int_{0}^{t}dt^{\prime}f(t^{\prime})\;I_{\mu,1-\lambda}(t-t^{\prime})+\;const^{\prime}e^{-\int_{0}^{t}R_{1-\lambda}(t^{\prime\prime})dt"}
\end{equation}
where 
\begin{equation}
I_{\mu,1-\lambda}\left(t\right)\equiv\int d\mu d\lambda\mu\left(1-\lambda\right)p\left(\lambda,\mu\right)e^{-\left(u+\omega\right)t}\ .
\end{equation}
For $P\left(\lambda\right)=const$: 
\begin{align}
I_{\mu,1-\lambda}\left(t\right) & =\int d\mu\mu P\left(\mu\right)e^{-\mu t}\int d\lambda\left(1-\lambda\right)Q\left(\lambda\right)e^{-\left(1-\lambda\right)t}=-\frac{e^{-t}(1-t)-1}{t^{2}}\int d\mu P(\mu)\mu e^{-\mu t}\nonumber \\
 & =\frac{1-e^{-t}(1-t)}{1-e^{-t}}\frac{1}{t}I_{\mu}(t)\sim\frac{1}{t}I_{\mu}(t)\nonumber \\
I_{\mu^{2}}\left(t\right) & =\frac{1-e^{-t}}{t}\int d\mu P(\mu)\mu^{2}e^{-\mu t}\simeq\frac{1}{t}\int d\mu P(\mu)\mu^{2}e^{-\mu t}\nonumber \\
I_{\left(1-\lambda\right)^{2}}\left(t\right) & =\frac{e^{-t}\left(2e^{t}-2t-t^{2}-2\right)}{t^{3}}\int d\mu P(\mu)e^{-\mu t}\simeq\frac{2}{t^{3}}\int d\mu P(\mu)e^{-\mu t}
\end{align}
As $f(t)\sim I_{\mu}$, we have that that $I_{\mu,1-\lambda}\left(t\right)$
falls faster than $f\left(t\right)$. This is also true for $I_{\mu^{2}}\left(t\right),I_{\left(1-\lambda\right)^{2}}\left(t\right)$.
Therefore 
\begin{equation}
R_{\mu,1-\lambda}\left(t\right)\sim e^{\int_{0}^{t}R_{1-\lambda}(t^{\prime\prime})dt"}f(t)\;\left[\int_{0}^{\infty}dt^{\prime}\;I_{\mu,1-\lambda}\left(t^{\prime}\right)\right]=R_{\mu}(t)\;\left[\int_{0}^{\infty}dt^{\prime}\;I_{\mu,1-\lambda}\left(t^{\prime}\right)\right]\ ,
\end{equation}
so $\left\langle \mu\left(1-\lambda\right)\right\rangle \sim R_{\mu}(t)\gg R_{\mu}(t)R_{1-\lambda}\left(t\right)\sim\left\langle \mu\right\rangle \left\langle 1-\lambda\right\rangle $,
and at large times $\left\langle \mu\left(1-\lambda\right)\right\rangle /\left\langle \mu\right\rangle \left\langle 1-\lambda\right\rangle \sim1/\left\langle \mu\right\rangle \sim\left\langle 1/\mu\right\rangle $,
where the last equality holds at least in the 3 cases considered.
To compute the correlation coefficient we use 
\begin{align}
R_{\mu^{2}}\left(t\right) & =R_{\mu}(t)\;\left[\int_{0}^{\infty}dt^{\prime}\;I_{\mu^{2}}\left(t^{\prime}\right)\right]\nonumber \\
R_{\left(1-\lambda\right)^{2}}\left(t\right) & =R_{\mu}(t)\;\left[\int_{0}^{\infty}dt^{\prime}\;I_{\left(1-\lambda\right)^{2}}\left(t^{\prime}\right)\right]\nonumber \\
\left\langle \mu^{2}\right\rangle -\left\langle \mu\right\rangle ^{2} & =R_{\mu}(t)\;\left[\int_{0}^{\infty}dt^{\prime}\;I_{\mu^{2}}\left(t^{\prime}\right)\right]-\left[R_{\mu}(t)\right]^{2}\left[\int_{0}^{\infty}dt^{\prime}\;I_{\mu}\left(t^{\prime}\right)\right]^{2}\sim R_{\mu^{2}}\left(t\right)\nonumber \\
\left\langle \left(1-\lambda\right)^{2}\right\rangle -\left\langle 1-\lambda\right\rangle ^{2} & \sim R_{\left(1-\lambda\right)^{2}}\left(t\right)
\end{align}
so the correlation coefficient 
\begin{align}
C_{1-\lambda,\mu}\left(t\right) & \equiv\frac{\left\langle \mu\left(1-\lambda\right)\right\rangle -\left\langle \mu\right\rangle \left\langle 1-\lambda\right\rangle }{\sqrt{\left[\left\langle \mu^{2}\right\rangle -\left\langle \mu\right\rangle ^{2}\right]\left[\left\langle \left(1-\lambda\right)^{2}\right\rangle -\left\langle 1-\lambda\right\rangle ^{2}\right]}}\nonumber \\
 & \sim\frac{R_{\mu,1-\lambda}\left(t\right)}{\sqrt{R_{\mu^{2}}\left(t\right)R_{\left(1-\lambda\right)^{2}}\left(t\right)}}\underset{t\rightarrow\infty}{\longrightarrow}\frac{\int_{0}^{\infty}dt^{\prime}\;I_{\mu,\left(1-\lambda\right)}\left(t^{\prime}\right)}{\sqrt{\left[\int_{0}^{\infty}dt^{\prime}\;I_{\mu^{2}}\left(t^{\prime}\right)\right]\left[\int_{0}^{\infty}dt^{\prime}\;I_{\left(1-\lambda\right)^{2}}\left(t^{\prime}\right)\right]}}\label{eq:C_tau_lambda}
\end{align}
As the integrands in the last expression are positive we have that
$C_{1-\lambda,\mu}(t\rightarrow\infty)\geq0$. Note that its exact
value depends on the entire distributions of $P\left(\mu\right),Q\left(\lambda\right)$,
while the long time behavior of $R_{\mu}(t)$, Eqs. (\ref{eq:R_tau_powerlaw},\ref{eq:R_tau_exp_Gauss}),
depends only on the tails of the distributions.

By keeping track of the corrections to the asymptotic value, one can
show that $\left\vert C_{1-\lambda,\mu}\left(t\rightarrow\infty\right)-C_{1-\lambda,\mu}\left(t\right)\right\vert $
approaches zero with the same long time behavior derived in the Aging
section above.

\subsection*{Appendix E: fixation probability in the presence of a cloud}

In this Appendix the derivation of Eq. (\ref{kimura}) is sketched.
More precisely, it is shown that if $\lambda_{b}=\lambda_{a}\left(1+\varepsilon\right)$
and $\mu_{b}=\mu_{a}\left(1+\varepsilon_{\mu}\right)$ with $\left|\varepsilon_{\mu}\right|,\left|\varepsilon\right|\ll1$,
then for $\frac{L_{a}}{L_{b}}<1$ the probability that a single $b$-mutant
will fix in the population is 
\[
\ln p_{a\rightarrow b}^{fix}=-\frac{M_{eff}^{2}}{M}\ln\frac{L_{a}}{L_{b}}
\]
up to subleading corrections, of order $\ln M$. For notational simplicity
we describe the case $\mu_{b}=\mu_{a}$$\equiv\mu$, and $\lambda_{b}=\lambda_{a}\left(1+\varepsilon\right)$,
the more general case with $\varepsilon_{\mu}\ne0$ follows from a
similar argument. Note that \emph{the cloud size is not assumed to
be negligible}. The cloud individuals are assumed to have negligible
self-reproduction, $\lambda_{c}=0$.

The stochastic process is described by $\left(n_{a},n_{b},n_{c}\right)$,
the number of individuals in states $a,b$ and the cloud respectively,
such that $n_{a}+n_{b},+n_{c}=M$. The reactions are 
\begin{align*}
 & \left(n_{a},n_{b},n_{c}\right)\underset{\lambda_{a}\frac{1}{M}n_{a}n_{b}}{\longrightarrow}\left(n_{a}+1,n_{b}-1,n_{c}\right)\ \ \ ;\ \ \ \left(n_{a},n_{b},n_{c}\right)\underset{\lambda_{b}\frac{1}{M}n_{a}n_{b}}{\longrightarrow}\left(n_{a}-1,n_{b}+1,n_{c}\right)\\
 & \left(n_{a},n_{b},n_{c}\right)\underset{\lambda_{a}\frac{1}{M}n_{a}n_{c}}{\longrightarrow}\left(n_{a}+1,n_{b},n_{c}-1\right)\ \ \ ;\ \ \ \left(n_{a},n_{b},n_{c}\right)\underset{\lambda_{b}\frac{1}{M}n_{b}n_{c}}{\longrightarrow}\left(n_{a},n_{b}+1,n_{c}-1\right)\\
 & \left(n_{a},n_{b},n_{c}\right)\underset{\mu_{a}n_{a}}{\longrightarrow}\left(n_{a}-1,n_{b},n_{c}+1\right)\ \ \ ;\ \ \ \left(n_{a},n_{b},n_{c}\right)\underset{\mu_{b}n_{b}}{\longrightarrow}\left(n_{a},n_{b}-1,n_{c}+1\right)
\end{align*}

Using standard large-deviation techniques {[}{[}refs{]}{]}, we need
to find the solution (`instanton') to Hamilton's equations with the
Hamiltonian

\begin{align*}
H & =\left[\left(e^{\hat{n}_{a}-\hat{n}_{b}}-1\right)\frac{\lambda_{a}}{M}+\left(e^{\hat{n}_{b}-\hat{n}_{a}}-1\right)\frac{\lambda_{b}}{M}\right]n_{a}n_{b}\\
 & +\left(e^{\hat{n}_{a}-\hat{n}_{c}}-1\right)n_{a}n_{c}\frac{\lambda_{a}}{M}+\mu\left(e^{\hat{n}_{c}-\hat{n}_{a}}-1\right)n_{a}\\
 & +\left(e^{\hat{n}_{b}-\hat{n}_{c}}-1\right)n_{b}n_{c}\frac{\lambda_{b}}{M}+\mu\left(e^{\hat{n}_{c}-\hat{n}_{b}}-1\right)n_{b}
\end{align*}
where $\left(n_{a},n_{b},n_{c}\right)$ are the canonical variables,
and $\left(\hat{n}_{a},\hat{n}_{b},\hat{n}_{c}\right)$ are the canonical
conjugates (momenta). Here we took $\mu_{a}=\mu_{b}=\mu$. The initial
conditions (at $t\rightarrow-\infty$): 
\[
n_{c}^{init}=M\mu/\lambda_{a}\ \ \ ;\ \ \ n_{1}^{init}=M\left(1-\mu/\lambda_{a}\right)\ \ \ ;\ \ \ n_{2}^{init}=0
\]
and $\hat{n}_{1,2,c}^{init}=0$, and the final conditions: 
\[
n_{c}^{final}=M\mu/\lambda_{a}\ \ \ ;\ \ \ n_{1}^{final}=0\ \ \ ;\ \ \ n_{2}^{final}=M\left(1-\mu/\lambda_{b}\right)
\]
The probability will be equal to $\ln p_{a\rightarrow b}^{fix}=S$,
where $S$ is the action of the path.

As the entire process takes time $O\left(1/\varepsilon\right)$, the
momenta will be $O\left(\varepsilon\right)$. As the changes in $\left\langle \lambda\right\rangle $
will be $O\left(\varepsilon\right)$, changes in $n_{c}$ will be
of the same order. Denote 
\begin{align*}
\hat{n}_{a,b,c}\left(t\right) & =\varepsilon\hat{\phi}_{a,b,c}\left(t\right)\\
n_{c}\left(t\right) & =n_{c}^{init}+\varepsilon f\left(t\right)
\end{align*}
where $\hat{\phi}_{a,b,c}$ and $f\left(t\right)$ will be order one
in $\varepsilon$, $O\left(\varepsilon^{0}\right)$.

Using this notation and expanding to lowest order non-zero in $\varepsilon$
the Hamiltonian reads 
\[
\frac{M}{\varepsilon^{2}\text{\ensuremath{\lambda_{a}}}}H=\left[-\hat{\phi}_{c}(n_{a}+n_{b})+n_{a}\hat{\phi}_{a}+n_{b}\hat{\phi}_{b}\right]f+n_{a}n_{b}(\hat{\phi}_{a}-\hat{\phi}_{b}-1)(\hat{\phi}_{a}-\hat{\phi}_{b})+n_{a}n_{c}^{init}(\hat{\phi}_{a}-\hat{\phi}_{c})^{2}+n_{b}n_{c}^{init}(\hat{\phi}_{b}-\hat{\phi}_{c})(\hat{\phi}_{b}-\hat{\phi}_{c}+1)
\]

\textbf{Fast-slow separation: }The instanton takes time $O\left(1/\varepsilon\right)$.
The equilibration with the cloud happens in time $O\left(1\right)$.
We therefore use a fast-slow separation, solving for $n_{c},\hat{\rho}_{c}$
at fixed $\rho_{1,2},\hat{\rho}_{1,2}$, and then solve for the slow
dynamics of $\rho_{1,2},\hat{\rho}_{1,2}$. Hamilton equations give
\begin{align*}
\frac{df}{dt} & =-\left(n_{a}+n_{b}\right)f-n_{c}^{init}\left(-2n_{b}\phi_{c}-2n_{a}\phi_{c}+2n_{b}\phi_{2}+2n_{a}\phi_{1}+n_{b}\right)\\
\frac{d\hat{\phi}_{c}}{dt} & =\left(n_{a}+n_{b}\right)\phi_{c}-n_{a}\phi_{1}-n_{b}\phi_{2}
\end{align*}
At fixed $\rho_{a,b},\phi_{a,b}$ the general solution with parameters
$c_{1},c_{2}$ 
\begin{align*}
f\left(t\right) & =\left(c_{1}-c_{2}n_{c}^{init}\right)e^{-\left(n_{a}+n_{b}\right)t}+c_{2}n_{c}^{init}e^{\left(n_{a}+n_{b}\right)t}-\frac{n_{b}n_{c}^{init}}{n_{a}+n_{b}}\\
\hat{\phi}_{c}\left(t\right) & =c_{2}e^{\left(n_{a}+n_{b}\right)t}+\frac{n_{a}\phi_{1}+n_{b}\phi_{2}}{n_{a}+n_{b}}
\end{align*}
for the solution not to diverge $c_{2}=0$. At times large compared
to $\left(n_{a}+n_{b}\right)^{-1}$, (note that this time scale is
$O\left(\varepsilon^{0}\right)$) they converge to 
\[
f^{fast}=-\frac{n_{b}n_{c}^{init}}{n_{a}+n_{b}}\ \ \ ;\ \ \ \hat{\phi}_{c}^{fast}=\frac{n_{a}\phi_{1}+n_{b}\phi_{2}}{n_{a}+n_{b}}
\]
Substituting these values into the Hamiltonian and setting $H=0$
we solve for $\hat{\phi}_{a}-\hat{\phi}_{b}$ and find 
\[
\hat{\phi}_{a}-\hat{\phi}_{b}=1+O\left(\varepsilon\right)\ .
\]
The rate of change in $\rho_{1,2}$will indeed be $O\left(1/\varepsilon\right)$$,$
as $\frac{dn_{a}}{dt}=\varepsilon\frac{n_{a}n_{b}\left(2\phi_{1}-2\phi_{2}-1\right)\left(n_{c}^{init}+n_{a}+n_{b}\right)}{n_{a}+n_{b}}$
and similarly for $n_{b}$, so time-scales separation is verified
self-consistently.

Using $\varepsilon\hat{\phi}_{1,2}=\hat{\rho}_{1,2}$, the action
reads 
\begin{align*}
S & =\int\hat{\rho}_{1}dn_{a}+\int\hat{\rho}_{2}dn_{b}=\int_{\rho_{1}^{init}}^{\rho_{1}^{final}}\left(\hat{\rho}_{1}-\hat{\rho}_{2}\right)dn_{a}\\
 & =\varepsilon M\left(1-\mu/\lambda_{a}\right)=M\left(1-\mu/\lambda_{a}\right)\ln\left(1+\varepsilon\right)=M_{eff}\ln\frac{\lambda_{b}}{\lambda_{a}}
\end{align*}
the last 2 equalities to $O\left(\varepsilon\right)$ as in the entire
calculation. The usual formula is restored with an effective population
size 
\[
M_{eff}=\rho_{1}^{init}=M\left(1-\mu/\lambda_{a}\right)\,\,,
\]
the size of the dominant population.

\emph{Genetic loads:} in terms of genetic loads, we find that here,
i.e. for $\mu_{1}=\mu_{2}=\mu$ and $\lambda_{b}=\lambda_{a}\left(1+\varepsilon\right)$
\begin{align*}
\frac{L_{a}}{L_{b}} & =\frac{\mu_{a}}{\mu_{b}}\frac{\lambda_{a}/\mu_{a}-1}{\lambda_{b}/\mu_{b}-1}=1-\varepsilon\frac{1}{1-\mu/\lambda_{a}}+O\left(\varepsilon^{2}\right)\\
\ln\frac{L_{b}}{L_{a}} & \simeq1-\frac{L_{a}}{L_{b}}\simeq\varepsilon\frac{1}{1-\mu/\lambda_{a}}=\varepsilon\frac{M}{M_{eff}}
\end{align*}
so with $M_{eff}=\rho_{1}^{init}=M\left(1-\mu/\lambda_{a}\right)$
\[
S=M_{eff}\ln\frac{\lambda_{b}}{\lambda_{a}}=M_{eff}\varepsilon=-\frac{M_{eff}^{2}}{M}\ln\frac{L_{a}}{L_{b}}
\]
as required.

\end{comment}
\end{thebibliography}


\begin{thebibliography}{10}
\bibitem{trap_model_comment}See \cite{Bouchaud}. The distibution
of $1/\mu$ in \cite{Bouchaud} has long tails; in the present work
we do not specialize on such distributions.
\end{thebibliography}
\end{document}